\documentclass[reqno]{amsart}

\usepackage{amssymb}
\usepackage{amsmath}
\usepackage{amsfonts,latexsym}
\usepackage{graphicx}

\newcommand{\bfalpha}{{\boldsymbol a}}
\newcommand{\supp}{{\text{supp}}}

\newcommand{\Gdt}{\begin{bmatrix}z^{-d}w(z)&-\phi(z)w(z)\\1&0\end{bmatrix}}

\newcommand{\hold}[1]{\mathcal{H}_{#1}}
\newcommand{\samp}[1]{\mathcal{S}_{#1}}

\newcommand{\Sn}{S_N}
\newcommand{\Comp}{\mathbb C}
\newcommand{\real}{{\mathbb R}}

\newcommand{\Z}{{\mathbb Z}}
\newcommand{\T}{\top}
\newcommand{\C}{\mathbf{C}}
\newcommand{\Hinf}{H^\infty}
\newcommand{\RHinf}{RH^\infty}

\newcommand{\Lt}{L^2}

\newcommand{\dd}{\mathrm{d}}
\newtheorem{prob}{\bf Problem}
\newtheorem{prop}{\bf Proposition}

\newtheorem{remark}{\bf Remark}
\newtheorem{coro}{\bf Corollary}

\begin{document}

\title[$H^\infty$ Optimal Approximation for Causal Spline Interpolation]{$H^\infty$ Optimal Approximation for Causal Spline Interpolation}
\author[M. Nagahara]{Masaaki Nagahara}
\author[Y. Yamamoto]{Yutaka Yamamoto}
\address{M. Nagahara and Y. Yamamoto are with 
Graduate School of Informatics, Kyoto University,
 Sakyo-ku Yoshida-Honmachi, Kyoto, 606-8501, Japan.
 The corresponding author is M. Nagahara (nagahara@ieee.org).}

\maketitle

\begin{abstract}
In this paper, we give a causal solution to the problem of 
spline interpolation using $H^\infty$ optimal approximation.  
Generally speaking, spline interpolation requires 
filtering
the whole sampled data, the past and the future, 
to reconstruct the inter-sample values.  
This leads to non-causality of the 
filter,
and this becomes a critical issue for real-time applications.  
Our objective here is to derive a causal system which approximates 
spline interpolation by $H^\infty$ optimization 
for the filter.
The advantage of $H^\infty$ optimization 
is that it can 
address uncertainty in the input signals to be interpolated in design,
and hence the optimized system has robustness property against signal uncertainty.
We give a 
closed-form
solution to the $\Hinf$ optimization 
in the case of the cubic splines.  
For 
higher-order
splines, the optimal filter can be effectively 
solved by a numerical computation.  
We also show that the optimal FIR
(Finite Impulse Response) filter can be designed
by an LMI (Linear Matrix Inequality),
which can also be effectively solved numerically.
A design example is presented to illustrate the result.
\end{abstract}


\section{Introduction}
\label{sec:intro}
Splines are widely used in image processing
due to their simple mathematical structure, 
in particular, linearity and low complexity
in computation.
Interpolation with these splines, called spline interpolation,
provides smoothness, that is,
the interpolated function can be continuous and 
several times differentiable.
By these advantages,
polynomial splines are very popular in image processing
such as curve fitting \cite{Sil85},
image interpolation (zooming) \cite{HouAnd78},
rotation \cite{UnsTheYar95},
compression \cite{DemDynIsk06},
and super resolution \cite{BabDra08}.

Theoretically, spline interpolation provides
perfect fitting for given sampled data
when the original analog signal is in the spline space
\cite{Sch67,Uns00,UnsAldEde93-1}.
This ideal spline interpolant
is 
however
obtained by 
filtering the whole sampled data.
This leads to {\it non-causality\/} of the interpolation process.
Although this non-causality is not a restriction for image processing,
spline interpolation cannot be used for 
real-time processing such as instrumentation or
audio/speech processing.
When
spline interpolation is used in AD (Analog-to-Digital) 
and DA (Digital-to-Analog) converters \cite{Uns99},
and when it is used in a feedback loop, the reconstruction delay degrades
the stability and the performance of the system.
In this case, the real-time processing is crucial.

For this non-causality problem, 
various approximation methods have been proposed
to obtain a {\it causal\/} system which approximates
the ideal (non-causal) spline interpolation,
by the constrained least square design \cite{UnsEde94},
the Kaiser window method \cite{VrcVai01},
and the maximum order minimum support (MOMS) function method
\cite{BluTheUns04}.
These methods are based on minimizing
the squared approximation error in the time domain.
This optimization can be generalized to $H^2$ optimization 
\cite{ZDG}.

$H^2$ optimization minimizes the $\ell^2$ norm of the impulse response.  
Hence it works basically for this particular signal only, and its 
performance against other input signals is not a priori guaranteed.
In other words, it can happen that the reconstruction error
will be significantly large for other unknown signals.
In real systems, input signals are unknown, or only partially known
(e.g., the input signals contain their frequencies mostly within 1.5 rad/sec),
and hence there are {\it uncertainty} in input signals.
To model such uncertainty, we
assume a certain class of input function spaces, and consider
a neighborhood (e.g., a unit ball) of such a function space.
We then consider that signal uncertainty as nominal signal plus
unknown signals that belongs to such a ball.  By controlling 
the induced norm of a pertinent operator, one can attenuate 
the response against such uncertainty, and this provides a
contrasting viewpoint of robustness, not in a probabilistic
sense, but in a deterministic treatment.  This has the 
advantage of minimizing the {\it worst-case\/} errors 
in contrast to probabilist models.
Robustness against such uncertainty is achieved by
{\it $H^\infty$ optimization} \cite{ZDG}
which
aims at minimizing or maintaining the error level below a 
certain prescribed performance level against all input signals, 
and possess much higher robustness
against lack of a priori knowledge about input signals to be processed.

$H^\infty$ optimization was first proposed and developed in control theory \cite{Zam81},
and then applied to signal processing \cite{CheFra95,YamAndNagKoy03,HasErdKai06}.
Since the $H^\infty$ norm gives the $\ell^2$-induced norm or the maximum energy gain,
minimizing the $H^\infty$ norm of the error system gives the optimality for the worst case.
This property leads to robustness of the system
against uncertainty of the input signal.
That is, the $H^\infty$ design
guarantees an error level $\gamma$ for all $\ell^2$ signals.
Moreover, the $H^\infty$ method can naturally take a frequency weight
in the design.
The weight can control the shape of the frequency response of the error system,
according to given knowledge on the frequency characteristic of the input signals.
From the computation viewpoint,
the $H^\infty$ optimization can be executed numerically via the state-space formulation,
and is easily done by standard softwares, as MATLAB.

We propose a new approximation method
for causal spline interpolation by $H^\infty$ optimization. 
The design is formulated as obtaining 
the $H^\infty$-sub-optimal
stable inverse filter of a system with unstable zeros.
In particular, for the cubic spline (3rd order spline),
the $H^\infty$ optimal filter can be obtained
in a closed form.
For a spline with arbitrary order, the $H^\infty$ 
sub-optimal
IIR (Infinite Impulse Response) filter
is easily obtained by numerical computation.
Moreover, by confining the desired 
sub-optimal filter to be FIR 
(Finite Impulse Response),
the optimization is reducible to an LMI (Linear Matrix Inequality),
which can be effectively solved by, for example,
standard MATLAB routines.
A design example is presented to show effectiveness of our method.

The paper is organized as follows.
In Section \ref{sec:spline-interpolation}, we introduce spline interpolation.
In Section \ref{sec:causal-spline}, we formulate our problem by $H^\infty$ optimization,
and derive the solution.
Performance analysis of our spline interpolation system is discussed in Section \ref{sec:analysis}.
Section \ref{sec:example} shows a design example and Section \ref{sec:conclusion} concludes our result.
\subsection*{Notation}
Throughout this paper, we use the following notation.
\begin{description}
\item[$\Z$, $\Z_+$]: the sets of integers and non-negative integers, respectively.
\item[$\real$, $\real_+$]: the sets of real numbers and non-negative
	real numbers, respectively.
\item[$\Comp$] the complex plane.
\item[${\mathcal D}$] the open unit disc in $\Comp$.
\item[$\C^N$] the set of continuous functions with continuous derivatives up to order $N$.
\item[$P_N$] the set of polynomial functions whose order is equal or less than $N$.
\item[$\Sn$] the set of $\C^N$ functions whose restriction on $[n,n+1)$, $n\in\Z$ is in $P_N$,
that is, on each interval $[n,n+1)$ the $\Sn$ function is a polynomial whose order is equal or less than $N$.
\item[$\Lt(X)$] the Lebesgue space consisting of all square integrable real functions on $X$. $\Lt(\real_+)$ is abbreviated to $\Lt$.
\item[$\ell^2(X)$] the set of all real-valued square summable sequences
	on $X$. $\ell^2(\Z_+)$ is abbreviated to $\ell^2$.
\item[$\delta$] the discrete-time impulse or the Kronecker delta, that is, $\delta(n)=1$, if $n=0$, and $0$, otherwise.
\item[$\phi\ast\psi$] convolution of a sequence $\{\phi(n)\}_{n\in\Z}$ and $\{\psi(n)\}_{n\in\Z}$, that is,
	\[
		(\phi\ast\psi)(n) = \sum_{n\in\Z} \phi(n-k)\ast\psi(k), \quad n\in\Z.
	\]
\item[$\sigma$, $\sigma^{-1}$] the forward and backward shift operator, respectively.
	That is, for a sequence $\{x(n)\}_{n\in \Z}$, $(\sigma \ast x)(n) = x(n+1)$ and $(\sigma^{-1} \ast x)(n) = x(n-1)$.
\item[$z$, $z^{-1}$] the $Z$-transform of $\sigma$ and $\sigma^{-1}$, respectively.
	For a sequence $\{x(n)\}_{n\in\Z}$, the $Z$-transform $\hat{x}$ of $x$
	is defined by
	\[
	 \hat{x}(z) := \sum_{n=-\infty}^{\infty} x(n)z^{-n}.
	\]
\item[$A^\T$] the transpose of a matrix $A$.
\item[$I_{M}$, $0_{M\times N}$] the $M\times M$ identity matrix and the
	   $M\times N$ zero matrix, respectively.
\end{description}
\section{Spline interpolation}
\label{sec:spline-interpolation}
We 
here discuss
polynomial spline interpolation.
In this paper, we consider the {\it cardinal interpolation problem} \cite{Sch67}:
\begin{prob}
Given a sequence $\{x(n)\}_{n\in \Z}$,
construct a function $y(t)$, $t\in\real$ satisfying the relation
\[
y(n) = x(n), \quad n\in\Z.
\]
\end{prob}
Needless to say, this problem is ill-posed because 
there are infinitely many solutions.  
To obtain a unique solution, one should specify the space to which 
the original signal $\{x(t)\}_{t\in \real}$ belongs.  
Assume that the space is
\[
V = \left\{ x\in \Lt(\real): \supp (\widehat{x}) \subseteq [-\pi,\pi] \right\},
\]
where $\Lt(\real)$ is the Lebesgue space of all square integrable functions on $\real$,
and $\widehat{x}$ is the Fourier transform of $x$.
Then, we have the well known solution called cardinal 
sinc
series \cite{Sch67,Uns00},
\[
y(t) = \sum_{n\in\Z} x(n) \frac{\sin \pi(t-n)}{\pi (t-n)},\quad t\in\real.
\]
That is, for any $x\in V$, we have $y(t)=x(t)$ for all $t\in\real$.

\begin{figure}[tb]
\begin{center}
\includegraphics[width=\linewidth]{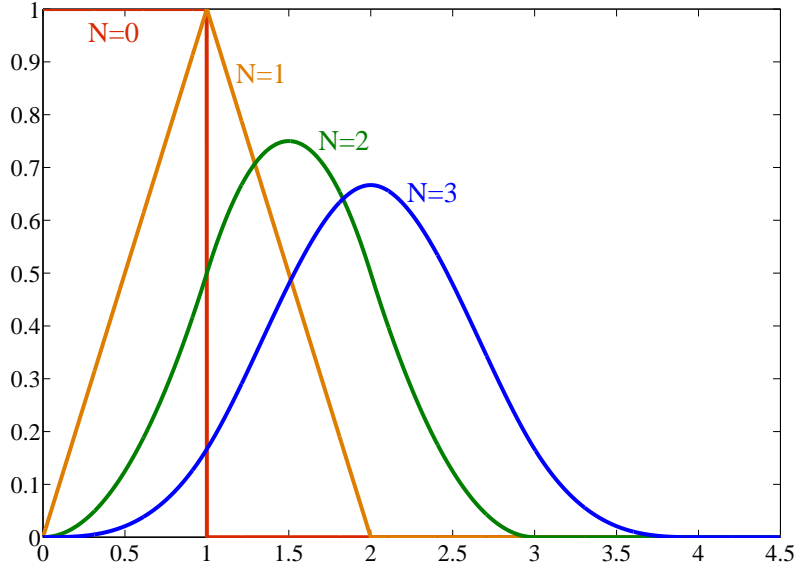}
\end{center}
\caption{Polynomial spline of order $N=0,1,2,3$}
\label{fig:spline}
\end{figure}

On the other hand,
assume that the space is
\begin{equation}
\Sn = \left\{x\in \C^N : x|_{[n,n+1)} \in P_N, n\in \Z\right\},
\label{eq:Sn}
\end{equation}
where $\C^N$ is the set of continuous functions with continuous derivatives up to order $N$.
Then the solution is given by 
\cite{Sch67},
\begin{equation}
\label{eq:b-spline}
y(t) = \sum_{n\in\Z} c(n) \phi(t-n), \quad t \in \real,
\end{equation}
where $\phi$ is the polynomial B-spline basis defined by \cite{Sch67,UnsAldEde93-1},
\[
\begin{split}
\phi(t) = (\underbrace{\beta^0 \ast \cdots \ast \beta^0}_{N+1})(t),\quad
\beta^0(t) = \begin{cases}1, & 0\leq t \leq 1, \\ 0, & \text{otherwise,}\end{cases}
\end{split}
\]
where `$\ast$' denotes convolution.  
Figure \ref{fig:spline} shows the polynomial splines $\phi(t)$ of order $N=0,1,2,3$.
In this formulation, 
the coefficients are given by the following convolution formula 
\cite{UnsAldEde93-1}: 
\begin{equation}
\label{eq:direct-transform}
c(n) = (\psi\ast x)(n),\quad n\in\Z,
\end{equation}
where $\psi$ is the direct B-spline filter satisfying 
$\psi\ast\phi=\delta$, 
or in $Z$-transform,
\begin{equation}
\label{eq:identity}
\psi(z)\phi(z)=1.
\end{equation}
This is for the perfect reconstruction without 
any delay. If we allow a delay $d>0$ for reconstruction,
the condition becomes $\psi\ast\phi=\sigma^{-d}$, 
where $\sigma^{-d}$ is the $d$-step delay, 
or the inverse $Z$-transform of $z^{-d}$,
that is,
\begin{equation}
\label{eq:identity-d}
\psi(z)\phi(z)=z^{-d}.
\end{equation}

\section{Causal spline interpolation by $\Hinf$ optimization}
\label{sec:causal-spline}
\subsection{Standard non-causal interpolation}
\begin{figure}[tb]
\begin{center}
\includegraphics[width=0.8\linewidth]{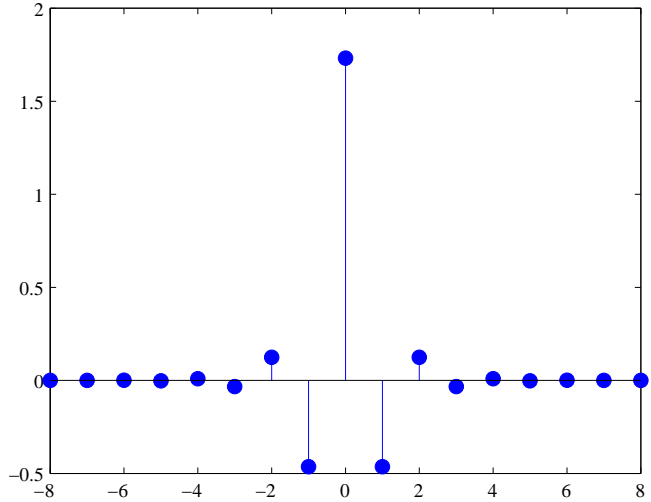}
\end{center}
\caption{Impulse response of non-causal filter $\psi(z)=6/(z+4+z^{-1})$}
\label{fig:spline_impulse}
\end{figure}

Since the $N$th-order spline $\phi(t)$ is supported in $[0,N+1)$,
the sampled signal $\phi(n)$
is represented as an FIR (finite impulse response) filter.
For example, in the case of $N=3$ (cubic spline),
we have
\begin{equation}
\label{eq:cubic-spline}
\phi(z) = \frac{1}{6}+\frac{2}{3}z^{-1}+\frac{1}{6}z^{-2}.
\end{equation}
By (\ref{eq:identity}), the desired filter $\psi(z)$ is given by 
the inverse $\psi=\phi^{-1}$ and it is seen that 
\[
\psi(z) = \frac{6}{z^{-2}+4z^{-1}+1}.
\]
One of the poles of $\psi(z)$ lies out of the open unit disc
$\mathcal{D}:=\{z\in\Comp: |z|<1\}$,
and hence the filter $\psi(z)$ becomes unstable.
The same can be said of the other $N$th-order splines \cite{UnsAldEde93-2}.
A practical way to implement this filter is 
to decompose $\psi(z)$ into a cascade of causal and anti-causal
filters \cite{UnsAldEde93-2}.
In the case of the cubic spline,
we first shift the impulse response of (\ref{eq:cubic-spline}) as
\[
z\phi(z) = \frac{1}{6}z+\frac{2}{3}+\frac{1}{6}z^{-1},
\]
and then decompose $\psi(z)=[z\phi(z)]^{-1}$ as
\begin{equation}
\psi(z) 
= -\frac{6\alpha}{1-\alpha^2}\left(\frac{1}{1-\alpha z^{-1}}+\frac{1}{1-\alpha z}-1\right),
\label{eq:psi-noncausal}
\end{equation}
where $\alpha = -2+\sqrt{3}$.
Since $|\alpha|<1$, this is a stable and {\it non-causal} IIR (infinite impulse response) filter.
Figure \ref{fig:spline_impulse} shows the impulse response of this non-causal filter.

\subsection{Causal interpolation by $\Hinf$ optimization}
\label{subsec:IIR}

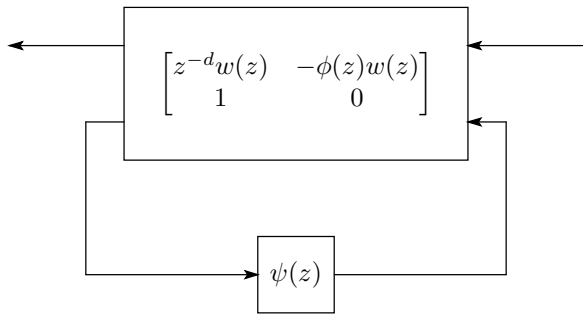
\begin{figure}[tb]
\begin{center}
\unitlength 0.1in
\begin{picture}( 30.0000, 16.0000)(  4.0000,-20.0000)
%
\special{pn 8}%
\special{pa 1000 400}%
\special{pa 2800 400}%
\special{pa 2800 1200}%
\special{pa 1000 1200}%
\special{pa 1000 400}%
\special{fp}%
%
\special{pn 8}%
\special{pa 1700 1600}%
\special{pa 2100 1600}%
\special{pa 2100 2000}%
\special{pa 1700 2000}%
\special{pa 1700 1600}%
\special{fp}%
%
\special{pn 8}%
\special{pa 1000 600}%
\special{pa 400 600}%
\special{fp}%
\special{sh 1}%
\special{pa 400 600}%
\special{pa 468 620}%
\special{pa 454 600}%
\special{pa 468 580}%
\special{pa 400 600}%
\special{fp}%
%
\special{pn 8}%
\special{pa 1000 1000}%
\special{pa 800 1000}%
\special{fp}%
%
\special{pn 8}%
\special{pa 800 1000}%
\special{pa 800 1800}%
\special{fp}%
%
\special{pn 8}%
\special{pa 800 1800}%
\special{pa 1700 1800}%
\special{fp}%
\special{sh 1}%
\special{pa 1700 1800}%
\special{pa 1634 1780}%
\special{pa 1648 1800}%
\special{pa 1634 1820}%
\special{pa 1700 1800}%
\special{fp}%
%
\special{pn 8}%
\special{pa 2100 1800}%
\special{pa 3000 1800}%
\special{fp}%
%
\special{pn 8}%
\special{pa 3000 1800}%
\special{pa 3000 1000}%
\special{fp}%
%
\special{pn 8}%
\special{pa 3000 1000}%
\special{pa 2800 1000}%
\special{fp}%
\special{sh 1}%
\special{pa 2800 1000}%
\special{pa 2868 1020}%
\special{pa 2854 1000}%
\special{pa 2868 980}%
\special{pa 2800 1000}%
\special{fp}%
%
\special{pn 8}%
\special{pa 3400 600}%
\special{pa 2800 600}%
\special{fp}%
\special{sh 1}%
\special{pa 2800 600}%
\special{pa 2868 620}%
\special{pa 2854 600}%
\special{pa 2868 580}%
\special{pa 2800 600}%
\special{fp}%
\put(19.0000,-8.0000){\makebox(0,0){$\Gdt$}}%
\put(19.0000,-18.0000){\makebox(0,0){$\psi(z)$}}%
\end{picture}%
\end{center}
\caption{Block diagram for $\Hinf$ optimization}
\label{fig:gplant_dt}
\end{figure}

In image processing, causality is often of secondary importance, 
and non-causal filters as above are used widely in that field, 
by suitably reversing the part of time axis as above.  
However, for real-time processing 
this is not quite appropriate, 
for example, in instrumentation or audio/speech processing.
To process such a signal, it takes infinite time or at least
time propotional to the length of the signal
since the non-causal filter $\psi(z)$ in (\ref{eq:psi-noncausal})
has infinite taps.
We propose a design of a causal filter $\psi(z)$ 
which approximates the condition (\ref{eq:identity-d}) of 
delayed perfect reconstruction, allowing a (small) 
time delay. 
Our problem is the following: 

\begin{prob}
Given a stable transfer function $\phi(z)$,
a stable weighting transfer function $w(z)$,
and delay $d\geq 0$,
find a causal and stable filter $\psi(z)$ which minimizes
\begin{equation}
\label{eq:J}
\begin{split}
J(\psi) 
&= \left\|\left\{z^{-d}-\psi(z)\phi(z)\right\}w(z)\right\|_\infty\\
&= \max_{\theta\in [0,2\pi)} 
\left|\left\{e^{-jd\theta}-\psi(e^{j\theta})\phi(e^{j\theta})\right\}w(e^{j\theta})\right|.
\end{split}
\end{equation}
\end{prob}
This is a standard $\Hinf$ optimization problem,
and it can be effectively solved by standard MATLAB routines
(e.g., \verb= dhfsyn = in MATLAB robust control toolbox \cite{BalChiPakSaf})
by using the block diagram shown in Figure \ref{fig:gplant_dt}. 
The MATLAB code for solving Problem 2 is available in \cite{web}.

\subsection{$\Hinf$ optimal cubic spline}

The cubic spline ($N=3$) is widely used because of its simple structure; 
for example, the cubic spline is the lowest-order spline for which the knot-discontinuity
is not visible to the human eye \cite{HasTibFri}.
Moreover, the cubic spline has
{\it minimum curvature property} \cite{Uns99},
that is, the cubic spline minimizes
\[
 \int_\real \left|y''(t)\right|^2 \dd t,
\]
the $L^2$ norm of the curvature of the interpolated signal $y(t)$
in (\ref{eq:b-spline}).
While the $H^{\infty}$ filter above can be effectively computed via 
various numerical methods, it is even possible to give a closed-form 
formula for the case of the cubic spline which is widely used
in digital signal processing.

Assume $w(z)=1$ and define
\[
E(z):=z^{-d}-\psi(z)\phi(z).
\]
Substituting (\ref{eq:cubic-spline}) into this equation, we have
\begin{gather*}
E(z) = z^{-d}-\psi(z)\frac{(z-\alpha_1)(z-\alpha_2)}{6z^2},\\
\alpha_1:=-2-\sqrt{3},\quad 
\alpha_2:=-2+\sqrt{3}.
\end{gather*}
This equation gives
\[
\psi(z) = \frac{6z^2(z^{-d}-E(z))}{(z-\alpha_1)(z-\alpha_2)}.
\]
Since $|\alpha_1|>1$, the filter $\psi(z)$ may have a pole 
outside the open unit disc ${\mathcal D}$.  
It is easily shown that the filter $\psi(z)$ is stable (i.e., all poles of $\psi(z)$ lie in $\mathcal{D}$) 
if and only if
\begin{equation}
\label{eq:interpolation}
E(\alpha_1) = \alpha_1^{-d}.
\end{equation}
Then our problem is to find a {\it stable\/}
$E(z)$ of minimum $\Hinf$ norm 
under the interpolation constraint (\ref{eq:interpolation}).
This is a {\it Nevanlinna-Pick interpolation problem\/} \cite{Wal}.
By the maximum modulus principle, we have
\[
\|E\|_\infty = \sup_{|z|=1}|E(z)| = \sup_{|z|\geq 1}|E(z)| \geq |E(\alpha_1)| = |\alpha_1^{-d}|.
\]
The interpolating function of minimum $H^\infty$ norm is therefore the constant function
$E(z) = \alpha_1^{-d}$.
By this, we obtain the optimal $\psi(z)$ as follows:
\begin{equation}
\label{eq:IIR3}
\begin{split}
\psi(z) &= \frac{6z^2}{(z-\alpha_1)(z-\alpha_2)}(z^{-d}-\alpha_1^{-d})\\
&= -\frac{6z^2}{\alpha_1^dz^d(z-\alpha_2)}\sum_{k=0}^{d-1}\alpha_1^{d-1-k}z^k.
\end{split}
\end{equation}

We summarize the result as a proposition.
\begin{prop}
For given $d\geq 0$ and the cubic spline function $\phi(z)$
in (\ref{eq:cubic-spline}),
the $H^\infty$ optimal $\psi(z)$ which minimizes
$J(\psi)=\|E\|_\infty$ is given by
\begin{equation}
\psi(z) = -\frac{6z^2}{\alpha_1^dz^d(z-\alpha_2)}\sum_{k=0}^{d-1}\alpha_1^{d-1-k}z^k,
\end{equation}
and the optimal value $\min_{\psi}J(\psi)=|\alpha_1^{-d}|$.
\end{prop}
\begin{remark}
For
higher-order
splines (i.e., $N\geq 4$), 
the optimal filter can be obtained by the 
{\it Nevanlinna algorithm} \cite{Wal}.  
A 
closed-form
solution is however very complicated
when $N\geq 4$.
In that case, the numerical computation
shown in \ref{subsec:IIR} or \ref{subsec:FIR}
is available.
\end{remark}

\subsection{FIR filter design via LMI}
\label{subsec:FIR}
The $\Hinf$-optimal filter is generally an IIR one.  
We here propose a design of the $H^\infty$-suboptimal FIR filter 
(with arbitrarily specified performance close to optimality).  
Assume that the direct filter $\psi(z)$ is FIR, that is,
\[
\psi(z) = \sum_{m=0}^M a_m z^{-m}.
\]
We here represent systems in a state space.
By the state-space formalism, we can reduce the
computation of $H^\infty$ optimization to
a linear matrix inequality (LMI).

A state-space representation of 
the FIR filter 
$\psi(z)$ is given by
\[
\begin{split}
\psi(z) &= \left[\begin{array}{ccccc|c}
	0     &1     &0     &\ldots&0     &0     \\
	\vdots&\ddots&\ddots&\ddots&\vdots&\vdots\\
	\vdots&      &\ddots&\ddots&0     &0     \\
	\vdots&      &      &\ddots&1     &0     \\
	0     &\ldots&\ldots&\ldots&0     &1     \\\hline
	a_M   &\ldots&\ldots&\ldots&a_1   &a_0
\end{array}\right](z)\\
&=: \left[\begin{array}{c|c}A_\psi&B_\psi\\\hline 
C_\psi(\bfalpha) & D_\psi(\bfalpha)\end{array}\right](z),
\end{split}
\]
where $\bfalpha := \left[\begin{array}{cccc}a_M&\ldots&a_1&a_0\end{array}\right]^\top$,
and we use the notation by Doyle \cite{ZDG}:
\[
\left[\begin{array}{c|c}
A&B\\\hline
C&D\\
\end{array}
\right](z)
:= C(zI-A)^{-1}B + D.
\]
Note that the parameter vector $\bfalpha$ to be designed is linearly
dependent 
only on the matrices $C_\psi(\bfalpha)$ and $D_\psi(\bfalpha)$,
that is, 
$C_\psi(\bfalpha) = \bfalpha^\top V_C$ and 
$D_\psi(\bfalpha) = \bfalpha^\top V_D$, 
where
\[
V_C =  \left[\begin{array}{c}I_M \\ 0_{1\times M}\end{array}\right],\quad
V_D =  \left[\begin{array}{c}0_{M\times 1} \\ 1\end{array}\right].
\]
Set state-space representations of $\phi(z)w(z)$ and $z^{-d}w(z)$ respectively by
\[
\begin{split}
\phi(z)w(z)&=:\left[\begin{array}{c|c}A_{\phi}&B_{\phi}\\\hline 
C_{\phi}& D_{\phi}\end{array}\right](z),\quad\\
z^{-d}w(z)&=:\left[\begin{array}{c|c}A_{d}&B_{d}\\\hline C_{d}& 0\end{array}\right](z).
\end{split}
\]
Then, a state-space representation of the error system
\[
E_w(z):= \left\{z^{-d}-\psi(z)\phi(z)\right\}w(z)
\]
is given by
\[
\begin{split}
E_w(z)
&= \left[\begin{array}{ccc|c}
	A_{\psi}&B_{\psi}C_{\phi}&0&-B_{\psi}D_{\phi}\\
	0&A_{\phi}&0&-B_{\phi}\\
	0&0&A_{d}&B_{d}\\\hline
	C_{\psi}(\bfalpha)&D_{\psi}(\bfalpha)C_{\phi}&C_{d}&-D_{\psi}(\bfalpha)D_{\phi}
\end{array}\right](z)\\
&=:\left[\begin{array}{c|c}A&B\\\hline C(\bfalpha)&D(\bfalpha)\end{array}\right](z).
\end{split}
\]
By this, the parameter $\bfalpha$ to be designed is affinely dependent
only on the matrices
$C(\bfalpha)$ and $D(\bfalpha)$, that is,
\[
\begin{split}
 C(\bfalpha) &= 
	\bfalpha^\top 
       \begin{bmatrix}V_C & V_D C_\phi & 0_{1\times d}\end{bmatrix}
	+ \begin{bmatrix}0_{1\times
		(M+N-1)}&C_d\end{bmatrix},\\
 D(\bfalpha) &= -\bfalpha^\top V_D D_\phi,
\end{split}
\]
where $N$ is the order of the B-spline basis $\phi$.
By using the bounded real lemma
or Kalman-Yakubovic-Popov (KYP) lemma,
we can describe our design problem as an LMI
\cite{YamAndNagKoy03}.
\begin{prop}
\label{prop:LMI}
Let $\gamma$ be a positive number. 
Then the inequality $\|E_w(z)\|_\infty<\gamma$ holds if and only if
there exist a positive definite matrix $P>0$ such that
\begin{equation}
\label{eq:LMI}
\left[\begin{array}{ccc}
	A^\T PA-P&A^\T PB&C(\bfalpha)^\T\\
	B^\T PA&-\gamma I+B^\T PB&D(\bfalpha)^\T\\
	C(\bfalpha)&D(\bfalpha)&-\gamma I
\end{array}\right]<0.
\end{equation}
\end{prop}
\begin{remark}
In some applications,
the error system $E_w(z)$ is required
to have specified zeros $z_i \in \Comp$,
$i=1, 2, \ldots, L$.
In particular, zero-bias constraint 
(i.e., $E_w(1)=0$ ) is used for perfect reconstruction
of DC (direct current) signals \cite{UnsEde94}.
In such cases,
zeros of $E_w(z)$ can be set by
\[
C(\bfalpha)(z_iI-A)^{-1}B + D(\bfalpha) =0,
\quad i=1, 2, \ldots, L.
\]
These are linear matrix equations
with respect to the design parameter $\bfalpha$.
The LMI (\ref{eq:LMI}) combined with 
these linear constraints is also easily solvable via
standard MATLAB routines. 
\end{remark}
\begin{remark}
To obtain the optimal $\bfalpha$,
minimize $\gamma$ subject to the LMI (\ref{eq:LMI}).
This minimization is also easily executed by MATLAB.
The MATLAB code for this optimization is available in \cite{web}.
\end{remark}
\section{Performance analysis}
\label{sec:analysis}
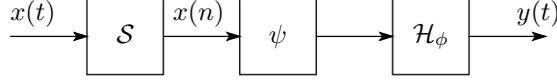
\begin{figure}[tb]
\begin{center}
\unitlength 0.1in
\begin{picture}( 28.0000,  4.2000)(  4.0000, -6.0000)
%
\special{pn 8}%
\special{pa 400 400}%
\special{pa 800 400}%
\special{fp}%
\special{sh 1}%
\special{pa 800 400}%
\special{pa 734 380}%
\special{pa 748 400}%
\special{pa 734 420}%
\special{pa 800 400}%
\special{fp}%
%
\special{pn 8}%
\special{pa 800 200}%
\special{pa 1200 200}%
\special{pa 1200 600}%
\special{pa 800 600}%
\special{pa 800 200}%
\special{fp}%
%
\special{pn 8}%
\special{pa 1200 400}%
\special{pa 1600 400}%
\special{fp}%
\special{sh 1}%
\special{pa 1600 400}%
\special{pa 1534 380}%
\special{pa 1548 400}%
\special{pa 1534 420}%
\special{pa 1600 400}%
\special{fp}%
%
\special{pn 8}%
\special{pa 1600 200}%
\special{pa 2000 200}%
\special{pa 2000 600}%
\special{pa 1600 600}%
\special{pa 1600 200}%
\special{fp}%
%
\special{pn 8}%
\special{pa 2000 400}%
\special{pa 2400 400}%
\special{fp}%
\special{sh 1}%
\special{pa 2400 400}%
\special{pa 2334 380}%
\special{pa 2348 400}%
\special{pa 2334 420}%
\special{pa 2400 400}%
\special{fp}%
%
\special{pn 8}%
\special{pa 2400 200}%
\special{pa 2800 200}%
\special{pa 2800 600}%
\special{pa 2400 600}%
\special{pa 2400 200}%
\special{fp}%
%
\special{pn 8}%
\special{pa 2800 400}%
\special{pa 3200 400}%
\special{fp}%
\special{sh 1}%
\special{pa 3200 400}%
\special{pa 3134 380}%
\special{pa 3148 400}%
\special{pa 3134 420}%
\special{pa 3200 400}%
\special{fp}%
\put(26.0000,-4.0000){\makebox(0,0){$\hold{\phi}$}}%
\put(18.0000,-4.0000){\makebox(0,0){$\psi$}}%
\put(10.0000,-4.0000){\makebox(0,0){$\samp{}$}}%
\put(4.0000,-3.5000){\makebox(0,0)[lb]{$x(t)$}}%
\put(12.5000,-3.5000){\makebox(0,0)[lb]{$x(n)$}}%
\put(30.5000,-3.5000){\makebox(0,0)[lb]{$y(t)$}}%
\end{picture}%
\end{center}
\caption{Signal Interpolation System}
\label{fig:reconstruction}
\end{figure}

In the previous section, we have proposed the $\Hinf$ optimization design of the filter $\phi(z)$
which approximates the delayed perfect reconstruction condition (\ref{eq:identity-d}).
In this section, we analyze the overall performance of the interpolation system shown in Figure \ref{fig:reconstruction}.
In Figure \ref{fig:reconstruction}, $\samp{}$ is the ideal sampler defined by
\[
\samp{}: \{x(t)\}_{t\in\real_+} \mapsto \{x(n)\}_{n\in\Z_+},
\]
and $\hold{\phi}$ is a hold
defined by
\[
\hold{\phi}: \{c(n)\}_{n\in\Z_+} \mapsto \left\{\sum_{n=0}^\infty c(n)\phi(t-n)\right\}_{t\in\real_+}.
\]
For simplicity, we set $w(z)=1$ in this section.
Then we show that the approximation of the equation (\ref{eq:identity-d}) is proper for decreasing 
the NSR (noise-to-signal ratio)
of the interpolation system.
\begin{prop}
\label{prop:performance-analog}
Assume that $\phi$ and $\psi$ are causal and stable.
Let $x$ be in $\Sn \cap \Lt$ and $y$ be the reconstructed signal 
by the direct B-spline transform $\psi$, that is,
\[
y(t) = \sum_{n=0}^\infty (\psi\ast x)(n) \phi(t-n), \quad t\in \real_+.
\]
Then there exists a real number $\lambda>0$ which depends only on $\phi$ such that 
for any non-negative integer $d$,
\[
\frac{\|x(\cdot-d)-y\|_{\Lt}}{\|x\|_{\Lt}} \leq \lambda J(\psi).
\]
\end{prop}
{\it Proof. }
Since $x\in\Sn$, there exists a sequence $\{c(n)\}_{n\in\Z_+}$
such that 
\[
x(t) = \sum_{n=0}^\infty c(n) \phi(t-n).
\]
We define $c(n)=0$ for $n<0$.
Then, for arbitrary fixed integer $d\geq 0$, we have
\[
\begin{split}
&x(t-d)-y(t)\\
&\quad = \sum_{n=0}^\infty \left\{c(n)\phi(t-d-n)-(\psi\ast x)(n)\phi(t-n)\right\}\\
&\quad = \sum_{n=0}^\infty \left\{c(n-d)-(\psi\ast x)(n)\right\}\phi(t-n)\\
&\quad = \sum_{n=0}^\infty \left\{c(n-d)-(\psi\ast\phi\ast c)(n)\right\}\phi(t-n)\\
&\quad = \sum_{n=0}^\infty \left\{(\sigma^{-d} - \psi\ast\phi)\ast c\right\}(n)\phi(t-n)\\
&\quad = \sum_{n=0}^\infty (e\ast c)(n) \phi(t-n),
\end{split}
\]
where $e:=\sigma^{-d}-\psi\ast\phi$.
Then, since $\phi$ is a Riesz basis \cite{StrNgu}, 
there exist $a>0$ and $b>0$ such that for any $c \in \ell^2$,
\[
a\|c\|_{\ell^2} \leq \left\|\sum_{n=0}^\infty c(n)\phi(t-n)\right\|_{\Lt} \leq b\|c\|_{\ell^2}.
\]
By using this inequality, we have
\[
\begin{split}
\|x(\cdot-d)-y\|_{\Lt} &= \left\|\sum_{n=0}^\infty (e\ast c)(n)\phi(\cdot -n)\right\|_{\Lt}\\
&\leq b \|e\ast c\|_{\ell^2}\\
&\leq b \|z^{-d}-\psi(z)\phi(z)\|_\infty \|c\|_{\ell^2}\\
&\leq \frac{b}{a} \|z^{-d}-\psi(z)\phi(z)\|_\infty \|x\|_{\Lt}.
\end{split}
\]
Since $J(\psi)=\|z^{-d}-\psi(z)\phi(z)\|_\infty$, we have
\[
\frac{\|x(\cdot-d)-y\|_{\Lt}}{\|x\|_{\Lt}} \leq \lambda J(\psi),
\]
where $\lambda=b/a>0$, which depends only on $\phi$.
\hfill$\Box$

We thus conclude that 
if the $\Hinf$ norm of the error system $z^{-d}-\phi(z)\psi(z)$
is adequately small, the NSR of the interpolator can be decreased,
and hence $\Hinf$ optimization provides a good approximation
of the ideal (i.e., non-causal) spline interpolation.

We next
consider a relation between our causal approximation and the ideal noncausal interpolation.
The following corollary to Proposition \ref{prop:performance-analog} guarantees that
our approximation recovers the ideal interpolation (i.e., perfect fitting)
when the delay $d$ goes to infinity.
\begin{coro}
Let $\RHinf$  be the set of all real, stable and causal IIR filters.
Then, for any $x\in\Sn\cap \Lt$ we have
\begin{equation}
\label{eq:prop2}
\inf_{\psi\in \RHinf} \frac{\|x(\cdot - d)-y\|_{\Lt}}{\|x\|_{\Lt}} \rightarrow 0,\quad \text{as } d \rightarrow \infty
\end{equation}
\end{coro}
{\it Proof. }
Let $J_{\text{opt}}(d)$ be the optimal value of $J(\psi)$, that is,
\[
J_{\text{opt}}(d) := \inf_{\psi\in \RHinf} \|z^{-d} - \psi(z)\phi(z)\|_\infty.
\]
Then we have \cite{CheFra95},
\[
\lim_{d\rightarrow\infty} J_{\text{opt}}(d) = 0.
\]
By this and Proposition \ref{prop:performance-analog}, we have (\ref{eq:prop2}).
\hfill$\Box$

The point of this proposition is that 
if we take sufficiently large delay $d$,
the worst-case approximation error is sufficiently small.

\section{Design Example}
\label{sec:example}
\begin{figure}[tb]
\begin{center}
\includegraphics[width=\linewidth]{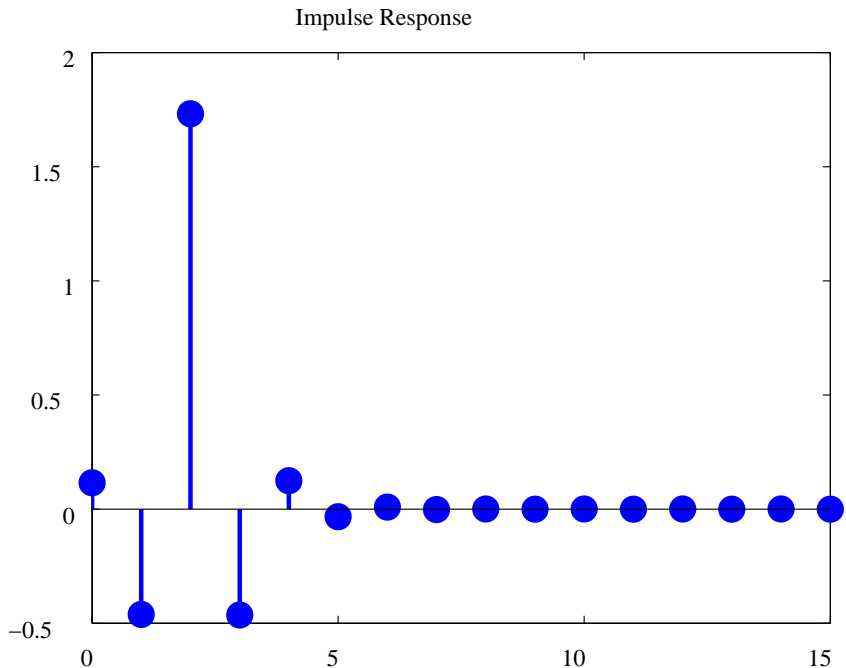}
\end{center}
\caption{Impulse response of $\Hinf$ optimal IIR filter $\psi(z)$}
\label{fig:psiHinf}
\end{figure}

\begin{figure}[tb]
\begin{center}
\includegraphics[width=\linewidth]{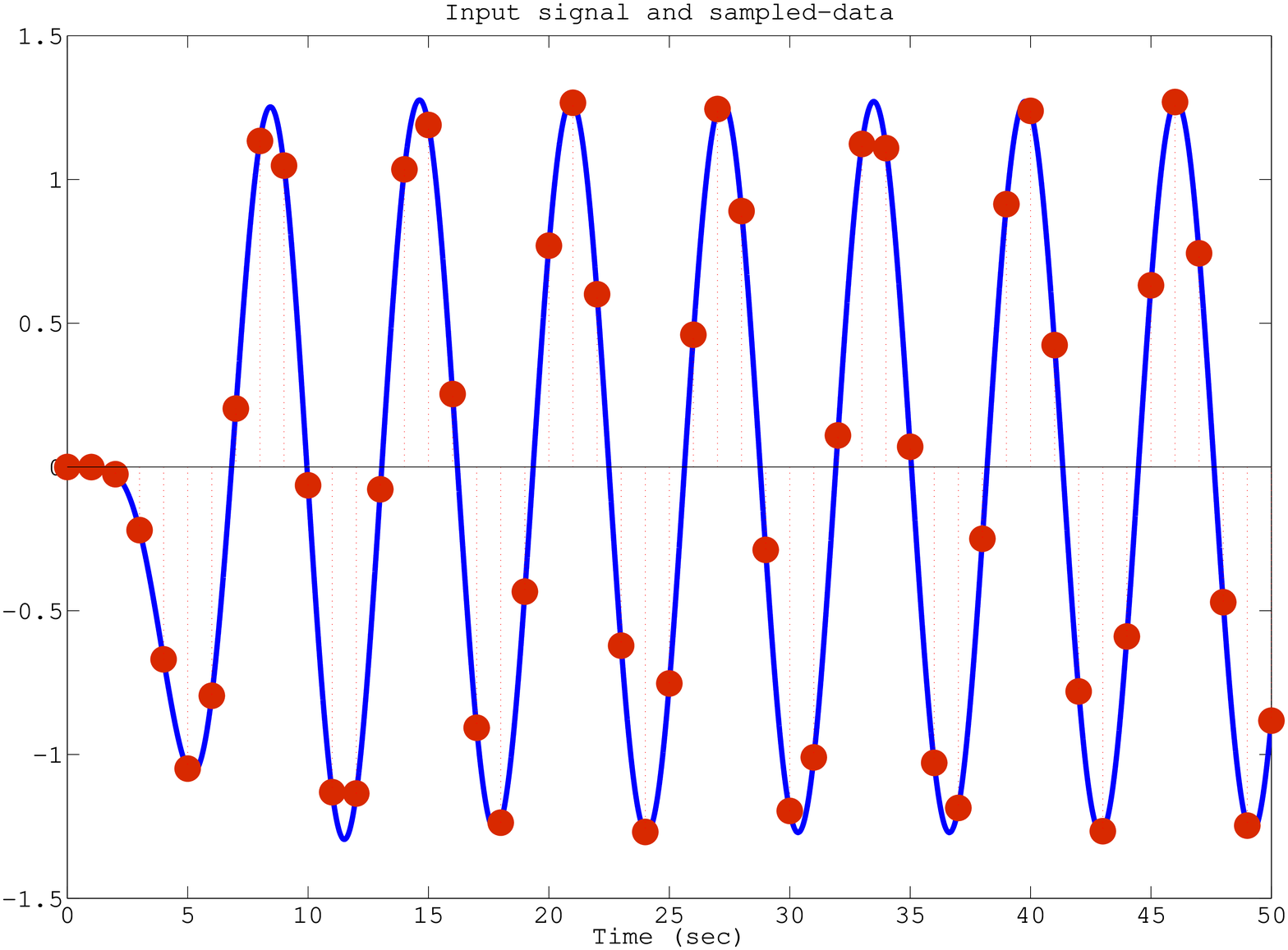}
\end{center}
\caption{The original analog signal and its sampled-data}
\label{fig:input}
\end{figure}

\begin{figure}[tb]
\begin{center}
\includegraphics[width=\linewidth]{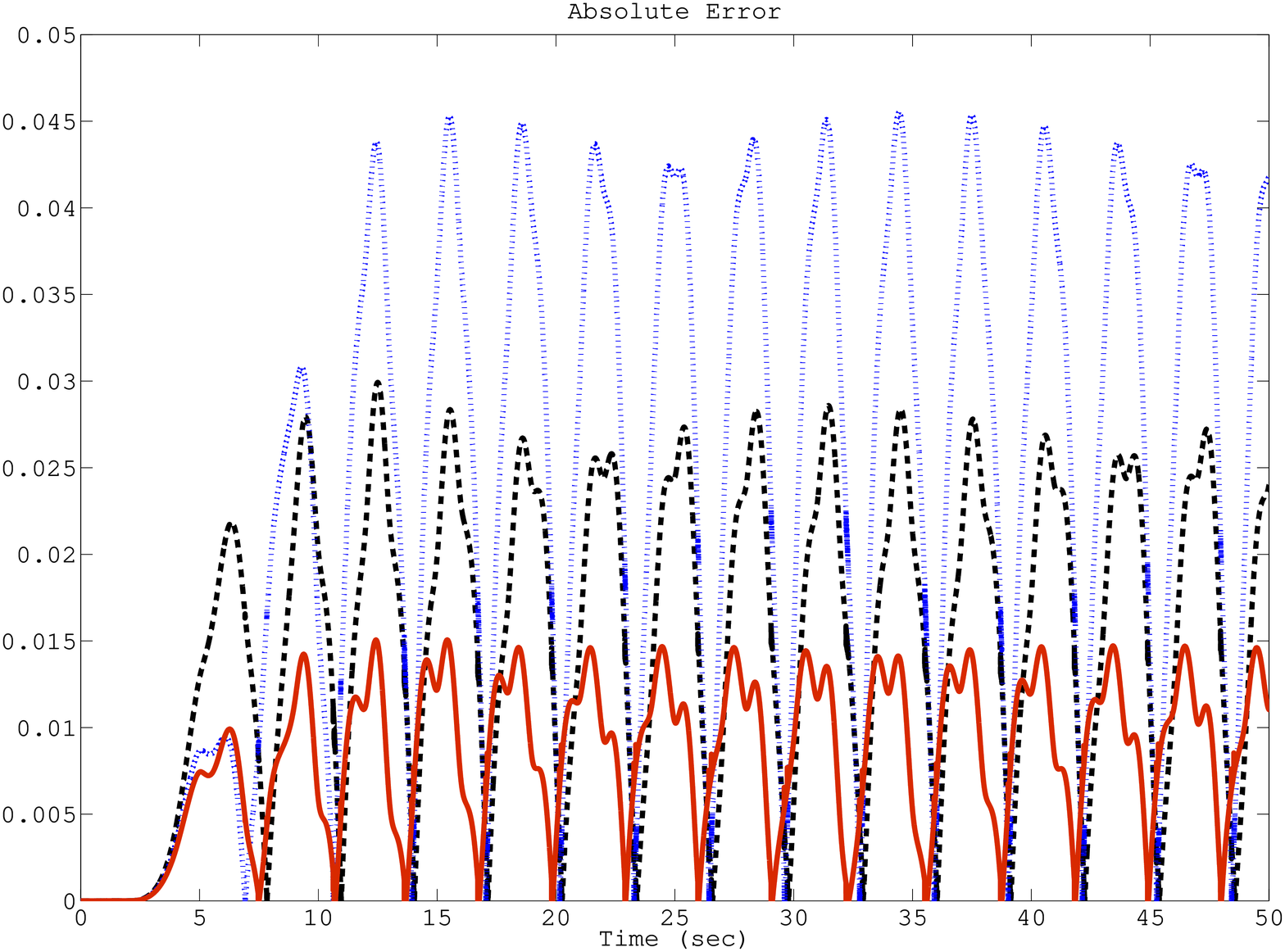}
\end{center}
\caption{Reconstruction error by weighted $\Hinf$ optimal FIR (solid),
unweighted $\Hinf$ optimal IIR (dash), and CLSD \cite{UnsEde94} (dots).}
\label{fig:error}
\end{figure}

We here present a design example of causal spline interpolation.
We consider the spline of order $N=3$ (cubic spline), 
take the reconstruction delay $d=3$,
assume $w(z)=1$,
and design the 
$\Hinf$ optimal IIR filter by (\ref{eq:IIR3}) 
and an FIR one with prespecified
degree of 5 taps using 
the linear matrix inequality (\ref{eq:LMI}).
In the case of the cubic spline, 
the $\Hinf$ optimal IIR filter (\ref{eq:IIR3}) 
with $d=3$ is given by
\begin{equation}
\psi(z) =  \frac{-6z^2-6\alpha_1 z-6\alpha_1^2}{\alpha_1^3z(z-\alpha_2)},
\label{eq:psiz}
\end{equation}
where $\alpha_1=-2-\sqrt{3}$ and $\alpha_2=-2+\sqrt{3}$.
Figure \ref{fig:psiHinf} shows the impulse response of this
filter.
For comparison, we also design a 5-tap FIR filter by
the constrained least square design (CLSD) \cite{UnsEde94}
and the Kaiser windowed approximation (KWA) \cite{VrcVai01}.
Table 1 shows the coefficients of the $\Hinf$ optimal
FIR filter, the filters by CLSD and by KWA.
\begin{table}[tb]
\begin{center}
{\bf Table 1}. Coefficient $a_k$ of FIR filter $\psi(z)$ \\
\begin{tabular}{|c|c|c|c|}\hline
$k$ & $\Hinf$ optimal & CLSD \cite{UnsEde94} & KWA \cite{VrcVai01} \\\hline
0 &  0.1152359 &   0.0991561 &        0.06049527\\
1 & -0.4614954 &  -0.4599156 &       -0.37739071\\
2 &  1.7307475 &   1.7215190 &        1.63379087\\
3 & -0.4614951 &  -0.4599156 &       -0.37739071\\
4 &  0.1152352 &   0.0991561 &        0.06049527\\\hline
\end{tabular}
\end{center}
\end{table}
\begin{figure}[tb]
\begin{center}
\includegraphics[width=\linewidth]{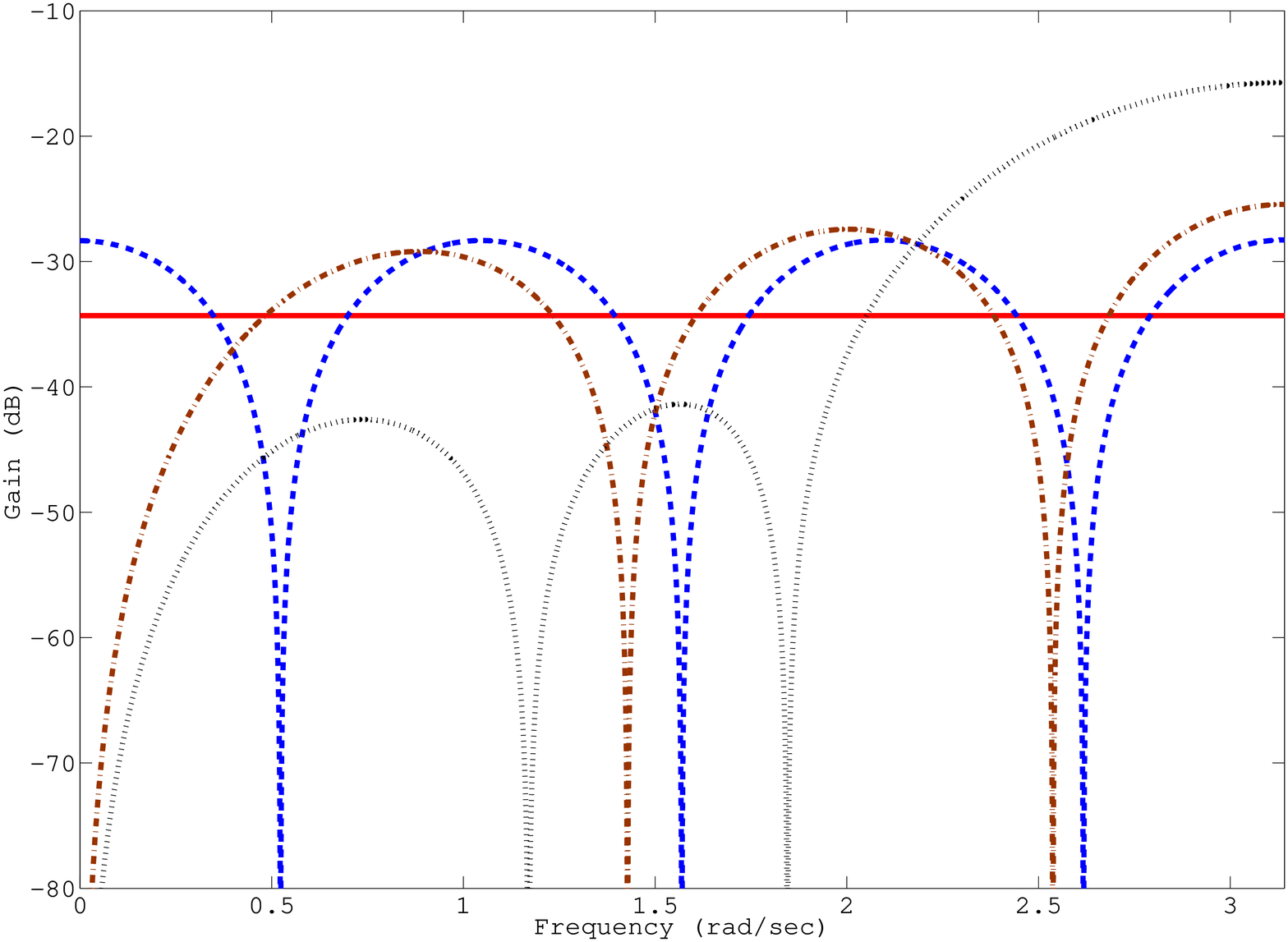}
\end{center}
\caption{Magnitude plot of $E(z)$: $\Hinf$ optimal IIR (solid),
$\Hinf$ optimal FIR (dash), CLSD \cite{UnsEde94} (dash-dots), and KWA \cite{VrcVai01} (dots).}
\label{fig:Ezmag}
\end{figure}
Figure \ref{fig:Ezmag} shows the magnitude of the frequency response
of the error system $E(z)=z^{-3}-\phi(z)\psi(z)$.
From this figure, we 
see that the $\Hinf$ optimal IIR filter given by (\ref{eq:psiz}) 
has the allpass characteristic.
The $\Hinf$ optimal FIR filter shows almost the same characteristic as
the CLSD filter except at the zero frequency.  
This is because CLSD aims at exact inversion for DC signals.
At the price of that, the CLSD filter exhibits 
larger errors in the high frequency range.
The KWA filter shows the same nature.  
Table 2 shows the $\Hinf$ norm of the error system $E(z)$.
\begin{table}[tb]
\begin{center}
{\bf Table 2}. $\Hinf$ norm of $E(z)$ \\
\begin{tabular}{|c|c|c|}\hline
Method & $\|E\|_\infty$ & $\|E\|_\infty$ in dB \\\hline
$\Hinf$ optimal IIR & 0.019238 & -34.3168\\
$\Hinf$ optimal FIR & 0.038597 & -28.2689\\
CLSD \cite{UnsEde94} & 0.053446 & -25.4417\\
KWA \cite{VrcVai01} & 0.16348 & -15.7307\\\hline
\end{tabular}
\end{center}
\end{table}
By Figure \ref{fig:Ezmag} and Table 2, we can see that the $H^\infty$ optimal
IIR filter is superior to CLSD by about 9 dB and KWA by about 19 dB
at the worst case frequencies.
In general, the purpose of $\Hinf$ design is to minimize the error 
in the worst case.  
This means that the $\Hinf$ design is 
against uncertainties in input signals, and this 
is an advantage of the $\Hinf$ design.  
While CLSD may perform better when 
the inputs can be predicted with 
certainty
(e.g., the inputs are all DC signals),
the $\Hinf$ design (worst case optimization) performs 
better when we do not have much information on 
the frequency characteristic of input signals. 
To see this robustness property of the $H^\infty$ method,
we simulate spline interpolation by $H^\infty$ method and CLSD.
The original analog signal is set to be 
the rectangular wave with the frequency 1 (rad/sec)
filtered by the 8-th order Butterworth lowpass filter
with the cut-off frequency 1.5 (rad/sec).
\begin{figure}[tb]
\begin{center}
\includegraphics[width=\linewidth]{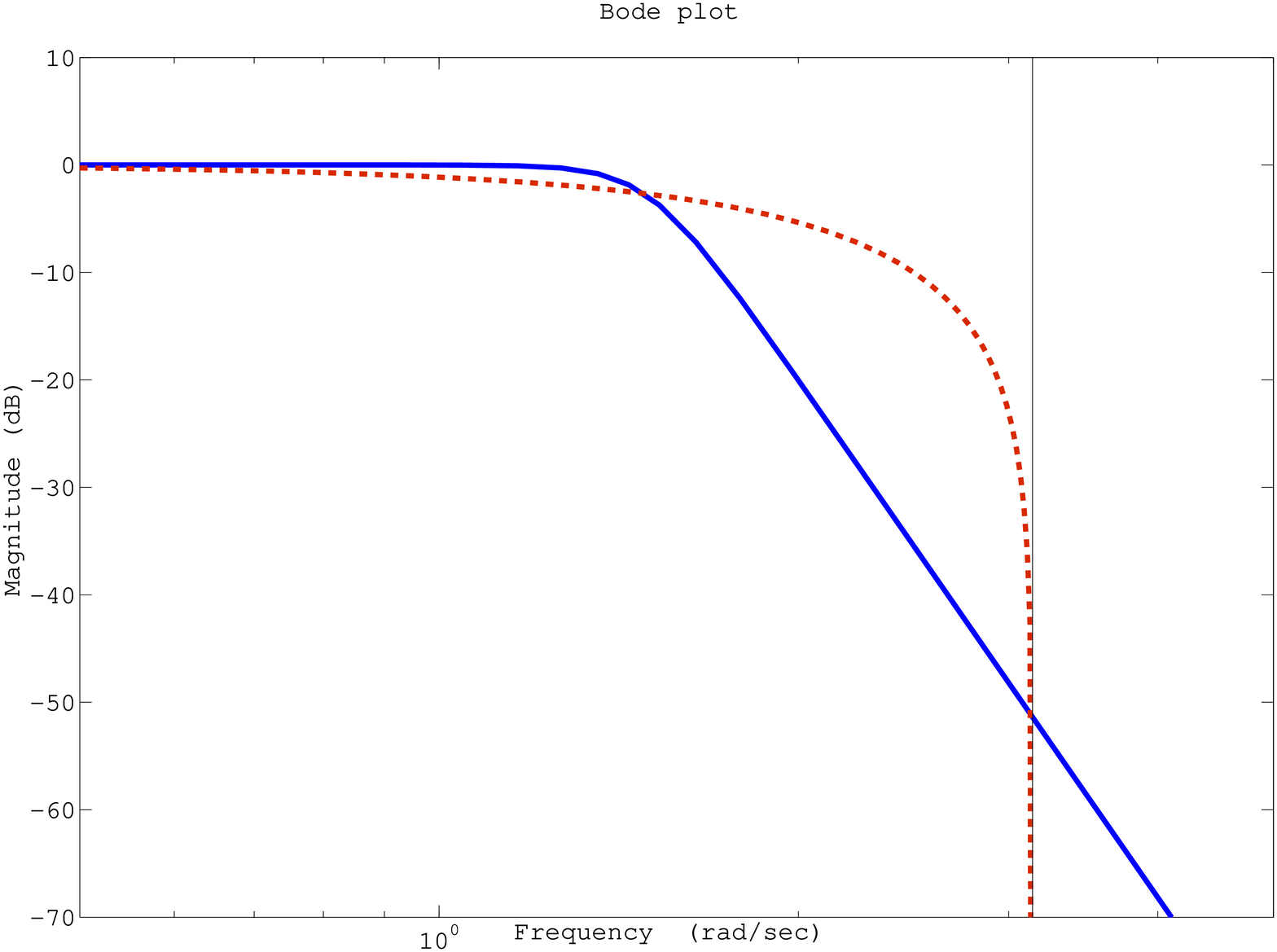}
\end{center}
\caption{Bode magnitude plot of 8-th order Butterworth lowpass filter (solid) and weighting function 
$w(z) = (1+z^{-1})/2$.}
\label{fig:weight}
\end{figure}
Figure \ref{fig:weight} shows the Bode magnitude plot of this lowpass filter,
and Figure \ref{fig:input} shows the analog signal filtered by the Butterworth filter and its sampled-data.
Note that this input signal is not exactly in the spline space $\Sn$ defined in (\ref{eq:Sn}).
This situation assumes that we have a priori
knowledge on the input analog signal
that the signal contains frequencies mostly in $[0,1.5]$ (rad/sec).
To bring this knowledge into our design,
we adopt the following frequency weight:
\[
 w(z) = \frac{1}{2}(1+z^{-1}).
\]
The Bode magnitude plot of $w(z)$ is shown in Figure \ref{fig:weight}.
With this weight, we design 5-tap FIR filter by the LMI in Proposition \ref{prop:LMI}.
Figure \ref{fig:error} shows the reconstruction errors
of the spline interpolation by this FIR filter, the $\Hinf$ optimal IIR filter given by (\ref{eq:psiz}),
and the CLSD filter given in Table 1.
Note that the errors in Figure \ref{fig:error} do not
vanish at the sampling instants since the original signal
$x(t)$ does not in the spline space $\Sn$.
The local minima in the errors are points at which
the original signal and the reconstructed one cross.
The $L^2$ norms of these errors are 1.9181 (CLSD),
1.2289 (unweighted $H^\infty$ optimal), and
0.7993 (weighted $H^\infty$ optimal).
The weighted $H^\infty$ optimal FIR filter shows the best performance
since this is designed with a priori knowledge on the signal frequency distribution.
On the other hand, the CLSD filter is designed to achieve perfect fit for DC signals,
but does not take other signals into account.
As a result it exhibits larger errors for unexpected signals as shown in Figure \ref{fig:Ezmag}.

There is also a design method for causal spline interpolation,
the maximum order minimum support (MOMS) function method
by Blu et al.~\cite{BluTheUns04}.
In contrast to the methods examined in this section,
the MOMS method optimizes the base functions.
To investigate robustness of the MOMS method by using the $H^\infty$
norm and to compare it with our method,
the optimality should be measured in sampled-data
$H^\infty$ norm \cite{CheFra}.
This is a theme for future study.

\section{Conclusion}
\label{sec:conclusion}
In this paper, we have proposed
a design of causal interpolation with
polynomial splines.
The design is formulated as an $\Hinf$ optimization problem.  
In the case of the cubic spline,
the optimal solution is given in a closed form.
Higher-order
optimal filters can effectively  be solved
by using MATLAB.
We have also shown that the $\Hinf$ optimal FIR filter
can be designed by an LMI.
A design example have been shown to illustrate the result.

A future topic is the $H^\infty$ design when $d$ is not
an integer, and also the order of the spline is fractional
\cite{UnsBlu00}.
This can be formulated by $H^\infty$ optimization for non-rational
transfer functions (or infinite-dimensional systems).




\begin{thebibliography}{99}
\bibitem{BabDra08}
L. Baboulaz and P. L. Dragotti,
Exact feature extraction using finite rate of innovation principles
with an application to image super-resolution,
{\it IEEE Trans.~Image Processing},
vol.~18,
no.~2,
pp.~281--298,
2009.
\bibitem{BalChiPakSaf}
G. Balas, R. Chiang, A. Packard and M. Safonov,
{\it Robust Control Toolbox Version 3},
The MathWorks,
2005.
\bibitem{BluTheUns04}
T. Blu, P. Th\'{e}venaz and M. Unser,
High-quality causal interpolation  for online unidimensional signal processing,
{\it Proc.~of the 12th EUSIPCO},
pp.~1417--1420,
2004.
\bibitem{CheFra}
T. Chen and B. A. Francis,
{\it Optimal Sampled-Data Control Systems},
Springer, 1995.
\bibitem{CheFra95}
T. Chen and B. A. Francis,
Design of multirate filter banks by $\mathcal{H}_\infty$ optimization,
{\it IEEE Trans.~Signal Processing},
vol.~43,
no.~12,
pp.~2822--2830,
1995.
\bibitem{DemDynIsk06}
L. Demaret, N. Dyn, and A. Iske,
Image compression by linear splines over adaptive triangulations,
{\it Signal Processing},
vol.~86,
pp.~1604--1616,
2006.
\bibitem{HasErdKai06}
B. Hassibi, A. T. Erdogan, and T. Kailath,
MIMO linear equalization with an $H^\infty$ criterion,
{\it IEEE Trans.~Signal Processing},
vol.~54,
no.~2,
pp.~499--511,
2006.
\bibitem{HasTibFri}
T. Hastie, R. Tibshirani, and J. Friedman,
{\it The Elements of Statistical Learning},
Springer, 2001.
\bibitem{HouAnd78}
H. Hou and H. C. Andrews,
Cubic splines for image interpolation and digital filtering,
{\it IEEE Trans.~Acoust., Speech, Signal Processing},
vol.~26,
no.~6,
pp.~508--517,
1978.
\bibitem{Sch67}
I. J. Schoenberg,
On spline interpolation at all integer points of the real axis,
{\it Delange-Pisot-Poitou.\ Theorie des nombres},
vol.~9,
no.~1,
pp.~1--18,
1967.
\bibitem{Sil85}
B. W. Silverman,
Some aspects of the spline smoothing approach to non-parametric
regression curve fitting,
{\it Journal of the Royal Statistical Society},
Series B,
vol.~47,
no.~1,
pp.\ 1--52,
1985.
\bibitem{StrNgu}
G. Strang and T. Nguyen,
{\it Wavelets and Filter Banks},
Wellesley-Cambridge Press,
1996.
\bibitem{Uns99}
M. Unser,
Splines: A perfect fit for signal and image processing,
{\it IEEE Signal Processing Magazine},
Vol.~16, No.~6, pp.~22--38, 1999.
\bibitem{Uns00}
M. Unser,
Sampling --- 50 years after {S}hannon,
{\it Proceedings of the IEEE},
vol.~88,
no.~4,
pp.~569--587,
2000.
\bibitem{UnsAldEde93-1}
M. Unser, A. Aldroubi and M. Eden,
B-Spline signal processing: Part-{I} --- Theory,
{\it IEEE Trans.~Signal Processing},
vol.~41,
no.~2,
pp.~821--833,
1993.
\bibitem{UnsAldEde93-2}
M. Unser, A. Aldroubi and M. Eden,
B-Spline signal processing: Part-{II} --- Efficient design and applications,
{\it IEEE Trans.~Signal Processing},
vol.~41,
no.~2,
pp.~834--848,
1993.
\bibitem{UnsBlu00}
M. Unser and T. Blu,
Fractional splines and wavelets,
{\it SIAM Rev.}, vol.~42, no.~1, pp.~43--67, 2000.
\bibitem{UnsEde94}
M. Unser and M. Eden,
{FIR} approximations of inverse filters and perfect reconstruction filter banks,
{\it Signal Processing},
vol.~36,
pp.~163--174,
1994.
\bibitem{UnsTheYar95}
M. Unser, P. Th\'{e}venaz, and L. Yaroslavsky,
Convolution-based interpolation for fast, high-quality rotation of
images,
{\it IEEE Trans.~Image Processing},
vol.~4,
no.~10,
pp.~1371--1381,
1995.
\bibitem{VrcVai01}
B. Vrcelj and P. P. Vaidyanathan,
Efficient implementation of all-digital interpolation,
{\it IEEE Trans.~Image Processing},
vol.~10,
no.~11,
pp.~1639--1646,
2001.
\bibitem{Wal}
J. L. Walsh,
{\it Interpolation and Approximation by Rational Functions in the
Complex Domain},
5th ed.,
{\it American Mathematical Society},
1969.
\bibitem{YamAndNagKoy03}
Y. Yamamoto, B. D. O. Anderson, M. Nagahara and Y. Koyanagi,
Optimizing {FIR} approximation for discrete-time {IIR} filters,
{\it IEEE Signal Processing Lett.},
vol.~10,
no.~9,
pp.~273--276,
2003.
\bibitem{Zam81}
G. Zames, 
Feedback and optimal sensitivity: model reference transformations, 
multiplicative seminorms and approximate inverses,
{\it IEEE Trans. Autom. Control}, 
vol.~26,
pp.~301--320,
1981.
\bibitem{ZDG}
K. Zhou, J. C. Doyle and K. Glover,
{\it Robust and Optimal Control},
Prentice Hall,
1995.
\bibitem{web}
{\tt http://www-ics.acs.i.kyoto-u.ac.jp/$^\sim$nagahara/cs/}
\end{thebibliography}

\end{document}